\documentclass[letterpaper, 10 pt, conference]{ieeeconf}  

\IEEEoverridecommandlockouts                              
\usepackage[utf8]{inputenc}
\usepackage[T1]{fontenc}
\usepackage{graphicx} 
\usepackage{xurl}

\title{\LARGE \bf
Human Biases Preventing The Widespread Adoption Of Self-Driving Cars
}

\author{Benjamin Kahl
\\
\today
}

\begin{document}

\maketitle
\thispagestyle{empty}
\pagestyle{empty}

\begin{abstract}

Self-driving cars offer a plethora of safety advantages over our accustomed human-driven ones, yet many individuals feel uneasy sharing the road with these machines and entrusting their lives to their driving capabilities.
Thus, bringing about a widespread adoption of autonomous cars requires overcoming these compulsions through careful planning and forethought. Here we break down the three primary psychological barriers that may hamstring or even wholly prevent their widespread adoption as well as how to tackle them.

\end{abstract}

\section{Introduction}

Recent strides in autonomous driving technology have brought self-driving cars ever closer to being regular, consumer-grade products. Their adoption promises to bestow improvements in road safety, efficiency and convenience.
But the biggest challenge to autonomous cars may not be technological, but psychological.

The vast majority of American citizens report feeling less safe at the prospect of sharing the road with self-driving vehicles and 78\% are afraid to ride in one. \cite{americans}

This simple fact may have far reaching implications. Anxious consumers may be susceptible to deterrence, provoke legislators into enacting suffocating regulations, even prompt some into outlawing them entirely.

The underlying apprehension is inextricably linked to the notion of \textit{trust}, and all its associated hopes and fears. Autonomous cars will be required to navigate complex traffic environments with the power of life and death over pedestrians, drivers and even their own passengers. Given this tremendous responsibility, consumers regard autonomous cars with trepidation and are unwilling to yield any vulnerability to a machine they don't trust.

Deconstructing this mistrust and fear as well as addressing it directly is quintessential towards paving the way for a widespread adoption of self-driving cars.

\section{Psychological Roadblocks to the adoption of self-driving cars}

Nature magazine article \textit{Psychological roadblocks to the adoption of self-driving vehicles} \cite{roadblocks} discusses three primary barriers impeding a widespread adoption and proposes steps towards addressing them: The dilemma of autonomous ethics, risk heuristics and the theory of the machine mind.

The subsequent chapters will present each as described by Shariff et al. in the \textit{Nature} article and provide further insight through data from a series of additional studies and surveys.

\begin{figure}[h]
    \centering
    \includegraphics[width=.48\textwidth]{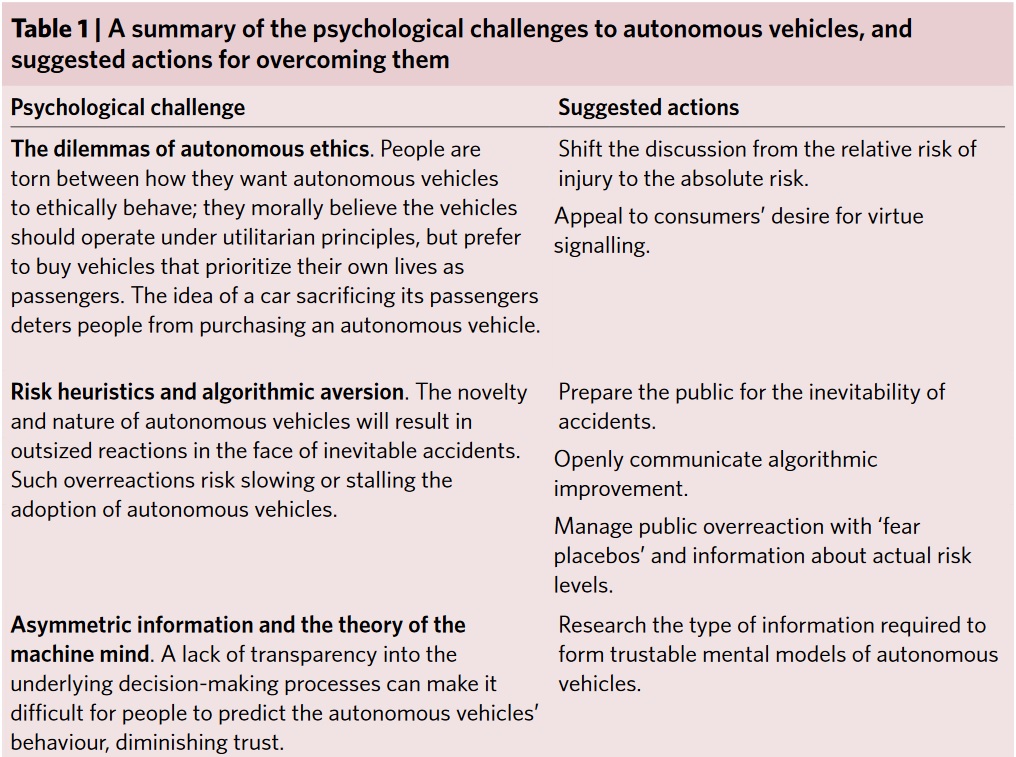}
    \caption{Overview of the primary psychological roadblocks, as shown in \textit{Psychological roadblocks to the adoption of self-driving vehicles} \cite{roadblocks}}
    \label{fig:rodablocks}
\end{figure}

\section{Dilemma of autonomous ethics}

Computer-driven cars are being heralded as the future benefactors of a new standard for road safety, in no small part thanks to their vitally short reaction-times and pragmatic decision making capabilities.

When faced with split-second decisions, human drivers cannot be expected to act in accordance of some written ruleset and thus have to rely on their neural precepts and natural instincts.

For this reason, the ability to assess and act upon difficult predicaments within a limited timespan has brought fourth a new set of moral conundrums to be grappled with.

Traditionally described in terms of a tram, railway and lever, \textit{the trolley problem} has found itself in the center of discussion about how autonomous cars should behave on the road, given that extreme traffic situations may lead to a choice of either putting the lives of several pedestrians into peril, or saving them by sacrificing its own passengers.

A cars modus operandi may adhere to a classical utilitarian doctrine by prioritizing the minimization of potential casualties in any given situation. Alternatively, the priority may lie in keeping the cars own passengers alive.

The \textit{dilemma of autonomous ethics} refers to the fact that potential self-driving vehicle purchasers are likelier to purchase self-protective models, yet want other consumers to drive utilitarian ones.

Potential utilitarian AV\footnote{AVs (autonomous vehicles) may include other vehicles than privately owned cars. For the sake of brevity the term \textit{AV} will be used synonymously to \textit{autonomous car} throughout the remainder of this document.}  buyers might be prompted to think twice when confronted with the thought that their car may one day decide to harm them.

\subsection{Potential counter-measures}

While government-imposed regulations may be required to enforce a utilitarian doctrine, they may also bring unintended side effects, as most consumers disapprove of such regulations \cite{dilemma}.

In addition, disallowing generally preferred self-protective vehicles may induce a delay in the widespread adoption of AVs, negating the potential lives saved by utilitarian models in the first place.

Shariff et al. assert that, in order to alleviate consumer deterrence, people need to feel both safe and virtuous about their AV ownership.

In order to invoke a sentiment of safety, it is imperative to keep the public informed on the overall safety benefits that self-driving vehicles can provide as well as highlighting the exceeding rarity of self-sacrifice scenarios.

To beseech a feeling of virtue, Shariff et al. suggest to mimic the appeal of virtue signalling, which has already found success in the marketing strategies of electric car manufacturers.

Vehicles such as the \textit{Toyota Prius} and the \textit{BMW i}-Series exhibit highly recognizable, unique shapes and color-schemes on their exterior designs in order to allow their owners to signal to the public their environmental commitment. Analogously, utilitarian AVs may allow them to signal their \textit{ethical} commitment.

\subsection{Social Dilemma of AVs}

This inherently social dilemma is further examined in the \textit{Science Journal} article \textit{The Social Dilemma of Autonomous Vehicles} \cite{dilemma} through a series of 6 surveys that closely emulate the \textit{Moral Machine Experiment} \cite{moralmachine}, although with a significantly smaller sample size and somewhat more specific inquiries.

\begin{figure}[h]
    \centering
    \includegraphics[width=.48\textwidth]{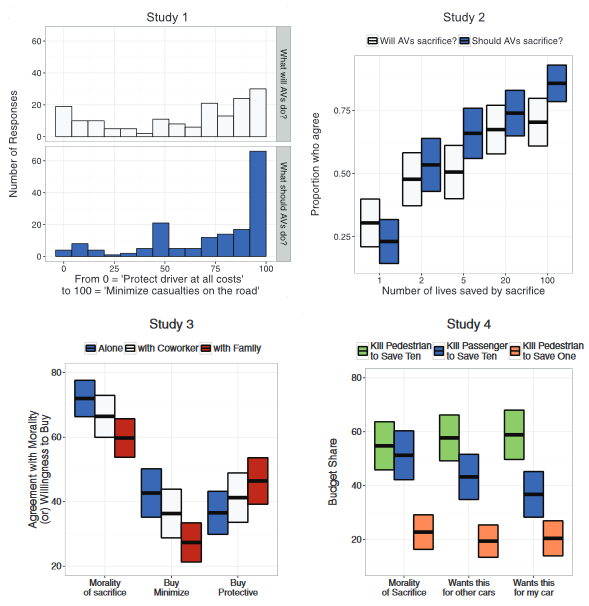}
    \caption{Results from studies 1-4 as presented in \textit{The social dilemma of autonomous vehicles}\cite{dilemma}}
    \label{fig:dilemma1}
\end{figure}

The results from the first and second study (see fig. \ref{fig:dilemma1}) clearly demonstrate the presence of overwhelming support for a utilitarian driving model, given the large preference for loss-minimizing choices. When confronted with the question of how AVs will actually behave, participants' responses were far more spread out, suggesting that AVs are believed to act in a less utilitarian manner than they should be.

The third study examines consumer preference for purchasing minimize or self-protective vehicles if the participant imagined themselves either alone, with a coworker or their family in the car. The results depicted in fig. \ref{fig:dilemma1} bring fourth following observations:

\begin{itemize}
    \item Despite overwhelming agreement with a loss-minimizing morality, the willingness to buy such a car is considerably lower.
    \item Agreement with loss-minimizing morality is palpably and negatively affected by the presence of a coworker or family member in the car.
    \item The willingness to purchase a self-protective vehicle increases in reverse proportion to the willingness of buying a loss-minimizing car.
\end{itemize}{}

Despite the willingness to purchase self-protective cars surpassing the willingness to buy minimize cars for multiple passengers, there is still a slight a preference for utilitarian ones if the participants imagine themselves alone.

This observation is of particular importance when considered in parity with the fact that average car occupancy rates lie well below 1.5 \cite{occupancy}, meaning most cars are driven alone.

Furthermore, an increased amount of passengers also dramatically decreases the likelihood of a self-sacrifice scenario, as an AV with 4 passengers would have to put the lives of \textit{at least} 4 pedestrians at risk in order to warrant such a maneuver.

\section{Risk heuristics}

Ever increasing adoption rates and higher amounts of AVs on the road will inevitably lead to to an increase in AV-related traffic accidents, despite their substantial safety advantages over human-driven cars.

However, due to their novelty, AVs will find a spotlight shone upon any traffic accident they are a part of (particularly lethal ones) in the form of disproportionately outsized media coverage.

Similarly to coverage on airplane crashes, AVs may find themselves stigmatized by the sheer amount of detrimental headlines and news articles which amplify peoples' fears by tapping into their

\begin{itemize}
    \item \textit{Availability bias}, which is the tendency to perceive examples of risks that come to mind easily as more representative than they genuinely are.
    \item \textit{Affective bias}, which causes risks to be perceived as larger if they evoke vivid emotional reactions.
    \item \textit{Algorithm aversion}, the tendency of people to loose faith in erroneous algorithms faster than in humans making similar mistakes.
    \item \textit{Overconfidence in ones own performance}, something highly prevalent in the driving domain in particular.
\end{itemize}{}

The exalted anxiety caused by negative media coverage paired with overconfidence in ones' own ability to stay out of accidents may compromise the financial feasibility of consumer-grade self-driving cars.

\subsection{Potential counter-measures}

Shariff et al. propose to prepare the public for the inevitability of AV accidents by not promising infallibility while still emphasizing the safety advantages AVs bring to the table.

Furthermore, legislators are to resist capitulating on peoples' fears and instead opt for the implementation of \textit{fear-placebos}. (High-visibility, low-cost concessions that do the most to quench public trepidation without posing a real obstacle to AV manufacturers.)

\subsection{Blame distribution in dual-error cases}

2016 brought this problem into the foreground with the first level 2 AV traffic fatality\footnote{As per the NHTSA\cite{nhsta}: A level 2 AV can itself control both steering and breaking/accelerating. The human driver must continue to pay full attention (“monitor the driving environment”) at all times and perform the rest of the driving task. It is considered to be a form of \textit{automated driving}, but not fully \textit{autonomous driving}.} in America, where a Tesla Model S engaged in \textit{Autopilot mode} crashed into an 18-wheel tractor-trailer killing its owner and only passenger.

The incident sparked a wave of news articles from every major news organization in the US, a level of coverage  that none of the over 40 thousand US traffic fatalities that year came even close to achieving.

In their official statement addressing the crash, Tesla claimed that \textit{"neither Autopilot nor the driver noticed the white side of the tractor trailer against a brightly lit sky, so the brake was not applied"}\cite{tesla_statement}, suggesting that both the human and machine driver should have taken action but neither did. Both drivers made a mistake.

Despite this, a considerable amount of the published news articles were remarkably skewed towards blaming the human driver.
Unsubstantiated rumors began circulating that the driver had been watching a movie\cite{harrypotter1}\cite{harrypotter2}\cite{harrypotter3} whilst on the road and had gotten multiple previous warnings\cite{warnings}.
Even the victims family themselves placed no blame on the car\cite{family}, despite having substantial grounds for a lawsuit\cite{lawsuit}.

The exact nature of blame distribution in dual-error cases between human and machine is examined in \textit{Blaming humans in autonomous vehicle accidents: Shared responsibility across levels of automation} by Awad et al. through a series of studies. 

The studies differentiate between a two types of drivers for each car: The \textit{primary driver} possesses general control over the vehicles' steering and acceleration at all times. If a mistake is made by the primary driver, it is the \textit{secondary drivers} obligation to intervene and prevent aforementioned mistake by overriding the primary drivers actions.

\begin{figure}[h]
    \centering
    \includegraphics[width=.48\textwidth]{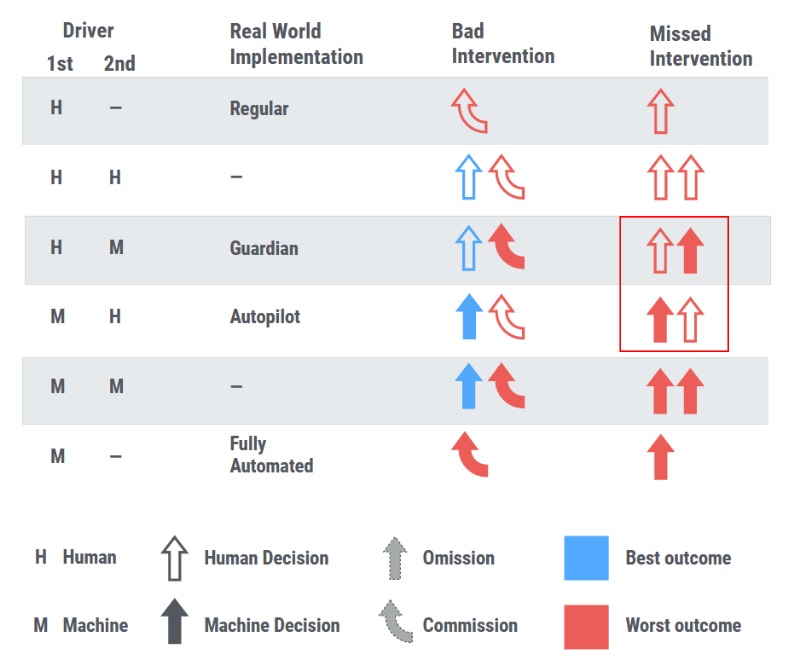}
    \caption{Scenarios regarded within the studies of \textit{Blaming humans in autonomous vehicle accidents: Shared responsibility across levels of automation}\cite{Blame}}
    \label{fig:blame_table}
\end{figure}

Figure \ref{fig:blame_table} lists all permutations of the regarded 2-driver scenarios, where H stands for human drivers and M for machine drivers. Thus, the MH scenario (Machine primary, human secondary) would correspond to the aforementioned Tesla Autopilot and the HM to the \textit{Toyota Guardian} system. Non critical cases (H, M, HH and MM) are merely taken for reference.

Fig. \ref{fig:blame_table} also shows the two types of regarded accidents:

\subsubsection{Bad Intervention}

In a \textit{bad intervention} accident the primary driver does not make a mistake. Instead, the secondary driver issues a faulty override which ultimately causes a crash.

As can be observed in fig. \ref{fig:blame_res}, the single mistake nature of this scenario leads to the vast majority of test participants placing the blame on the secondary driver.

\subsubsection{Missed Intervention}

A \textit{missed intervention} accident follows the pattern of the above mentioned 2016 Tesla crash: The primary driver makes a mistake, putting the car in peril. However the secondary driver fails to identify said mistake and thus does not intervene.

\begin{figure}[h]
    \centering
    \includegraphics[width=.5\textwidth]{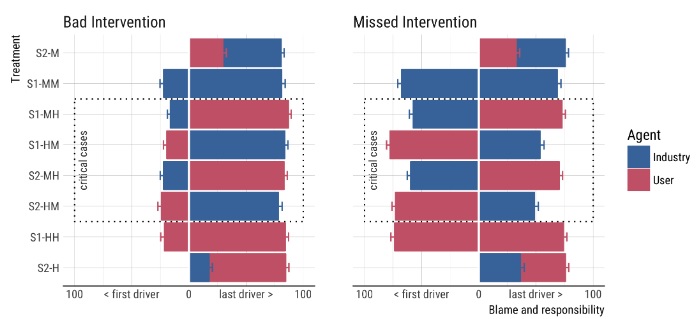}
    \caption{Results from the studies conducted in \textit{Blaming humans in autonomous vehicle accidents: Shared responsibility across levels of automation}\cite{Blame}. S1 and S2 refer to different studies.}
    \label{fig:blame_res}
\end{figure}

The results shown in fig. \ref{fig:blame_res} put on display that, no matter the type of accident and the allocation of roles (HM/MH), participants had a clear tendency to place more blame on the machine driver than the human driver. The \textit{man-in-the-loop} is blamed more than the \textit{machine-in-the-loop}.

Awad et al. conclude that, \textit{"while there may be many psychological barriers to self-driving car adoption, public over-reaction to dual-error cases is not likely to be one of them. In fact, we should perhaps be concerned about public under-reaction"}.

While bad intervention cases may be causal to the underlying psychological roadblock, the diminished blame on AV manufacturers in missed intervention cases may bring a different set of problems with it. It may, for instance, prompt manufacturers to design their Guardian systems more passively, as to avoid bad intervention failures and allocating traffic accidents caused by their systems further towards the missed-intervention spectrum.

In addition to the lesser pressure on manufacturers to perform, the foregoing disbalance may induce societal level responses to this trend. With the majority of the blame being allocated on the human, people may train themselves to drive so that, in the case of a crash, the blame would befall on the machine. (By, for example, not attempting to correct a faulty bad intervention.)

\section{Lack of transparency}

The final roadblock presented by Shariff et al. is the lack of transparency into the machine decision making process and the lack of trust inherently caused by it.

A central part of this barrier is the information asymmetry of consumers being highly aware of the individual flaws and failures within the AV domain, yet blissfully unaware of the plethora of small successes and optimizations.

Partially fuelled by the black-box nature of machine learning algorithms, the scarcity of exact knowledge into how AVs make their decisions turn them into unpredictable agents. This causes not just potential customers but also other road participants to approach AVs with some trepidation in regular traffic environments.

On the other hand, providing the public with a full disclosure of information about how AVs work internally would simply overwhelm the average consumer with the tremendous amount of technical details and intricacies that go into developing a self-driving vehicle.

The solution proposed by Shariff et al. is to conduct specialized research into what type of information a consumer should be exposed to in order to form trustworthy bonds between humans and machines, and what information should remain obscured through a series of abstractions and simplifications.

To some extent, the lack of transparency and predictability of AV decision making may also be linked to the disproportionate amount of attention placed of fringe moral scenarios, such as the trolley problem.

\subsection{Moral preferences for AVs}

The data portrayed in fig. \ref{fig:dilemma1} (Roadblock 1) seems to suggest that people have a general preference for strictly utilitarian driving models, but here participants were confronted with classical trolley-problem dilemmas where the choice is binary and the outcome probability is absolute.

In the publication \textit{How Should Autonomous Cars Drive? A Preference for Defaults in Moral Judgments Under Risk and Uncertainty}\cite{how} Meder at al. claim that the abstract nature of such dilemmas are highly unrealistic and thus inadequate for the purpose of modelling how AVs should drive.

The trolley problem is typically presented as a choice between \textit{action} (pulling a lever to prevent a larger number of deaths but actively cause a different set of people to die) or \textit{inaction} (abstain from any action that may cause harm to someone but allow a greater number of people to be harmed). In the domain of AVs, the problem can be modelled as either staying in lane and harming a pedestrian that is crossing the street or swerving to the side and possibly crash into a bystander standing on the sidewalk. 

Neither human-driven nor computer-controlled cars can possibly assess the exact outcome of any situation with flawless precision, thus these decisions have to be made on a probabilistic basis at best. In their paper, Meder et al. back this point up by quoting an engineer working on AVs:\\

\textit{"It takes some of the intellectual intrigue out of the [trolley] problem, but the answer is almost always “slam on the brakes”.... You’re much more conﬁdent about things directly in front of you, just because of how the system works, but also your control is much more precise by slamming on the brakes than trying to swerve into anything." (Hern, 2016)}\\

This emphasizes a higher degree of uncertainty when certain actions (swerving) are taken in contrast to others (breaking). In order to discover how participants believe AVs should behave under such circumstances, the outcomes of the choices offered in these surveys were only probabilistically known or entirely unknown.

The study is divided into a set of two experiments:\\

\subsubsection{Experiment 1}

\begin{figure}[h]
    \centering
    \includegraphics[width=.4\textwidth]{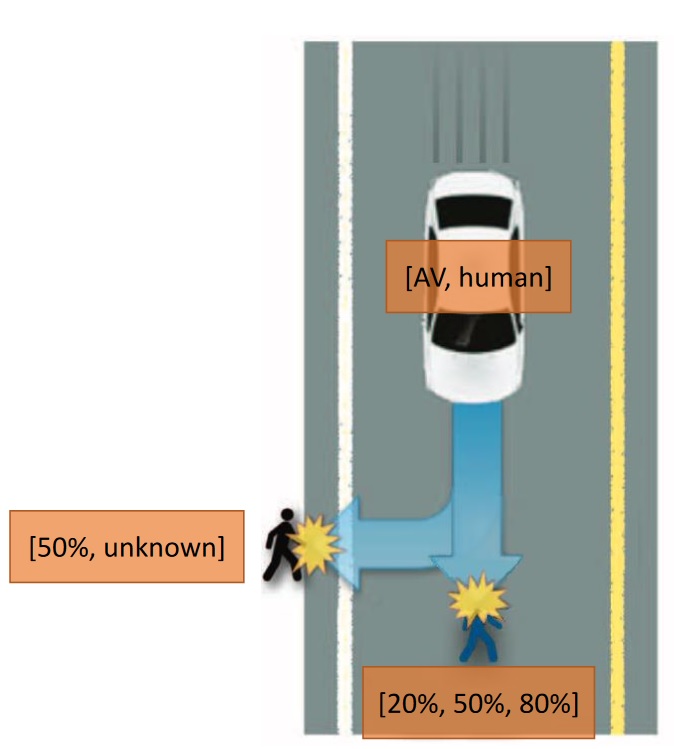}
    \caption{Predicament described in the first experiment of \textit{How Should Autonomous Cars Drive? A Preference for Defaults in Moral Judgments Under Risk and Uncertainty}\cite{how}. Depiction adapted from \textit{The social dilemma of autonomous vehicles}\cite{dilemma}}
    \label{fig:how_setup}
\end{figure}

The predicament described to participants in experiment 1 follows the depiction in fig. \ref{fig:how_setup}:

A car (either human-driven or AV) is moving along a street when it suddenly encounters a pedestrian on the street in front of it. The chances of collision are assessed by the driver as either 20\%, 50\% or 80\%.

The driver must now choose to either put on the brakes while staying in lane and cause a lethal accident with the above mentioned probability, or alternatively, it can try to swerve to the side and negate the chances of colliding with the pedestrian entirely. Instead, if the car swerves, there is either a 50\% or an unknown chance to collide with an unrelated bystander.

The survey was taken by a total of 1648 participants of which 872 were used in result analysis, as the remaining 776 either dropping out or were excluded on grounds of insufficient language proficiency or previous participation.

Each test subject was asked a series of 5 questions:
\begin{itemize}
    \item \textbf{Decision Preference:} Should the car stay or swerve?
    \item \textbf{Moral Judgement:} How morally acceptable is it to stay or swerve?
    \item \textbf{Swerving Threshold:} Participants were asked to provide a minimum likelihood of collision at which to swerve.
    \item \textbf{Probability Estimate:} If the probability of collision with the bystander was unknown, participants were asked to provide an estimate.
    \item \textbf{Decision Rule:} What general rule should drivers follow in such situations?
\end{itemize}{}

\begin{figure}[h]
    \centering
    \includegraphics[width=.48\textwidth]{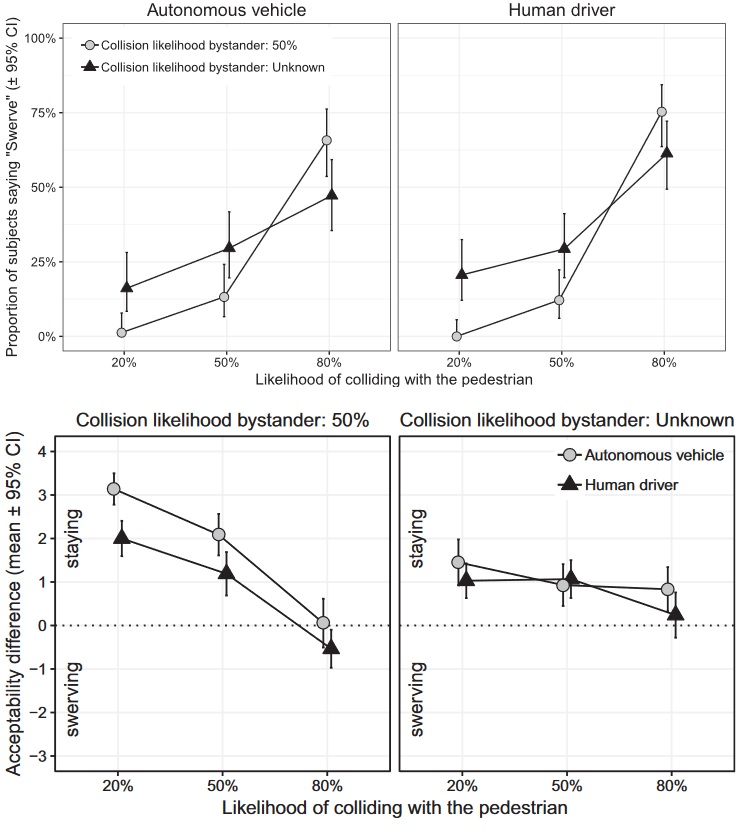}
    \caption{Results to first and second questions of experiment 1.}
    \label{fig:how_q1_q2}
\end{figure}

Fig. \ref{fig:how_q1_q2} shows the distribution of answers given to the first question. As one may expect, the proportion of people opting for the "swerve" option increases in relation to the chances of collision with the pedestrian.

However, rudimentary utilitarian arithmetic would yield a graph passing through the 50\%-50\% intersection, as the chances of killing either the pedestrian or bystander are equal. Instead, participants showed a higher preference for staying in lane and thus, evidently, did not adhere to a simple consequentialist analysis.

Equally noteworthy is the fact that, if the chances of collision with bystander were unknown, participants preferred the car to stay in lane, even if likelihood of collision with pedestrian were unusually high (80\%). In addition, this only holds true for AVs and not human drivers, meaning that, according to participants, AVs should be less prone to swerving maneuvers than their human counterparts.

A similar trend can be observed in the results of the second question, with almost all data-points hovering on the side of staying. Furthermore, most AV data-points are above those of human drivers, meaning that participants found it to be \textit{more moral} for computer-driven cars to stay in lane than human-driven ones.

\begin{figure}[h]
    \centering
    \includegraphics[width=.48\textwidth]{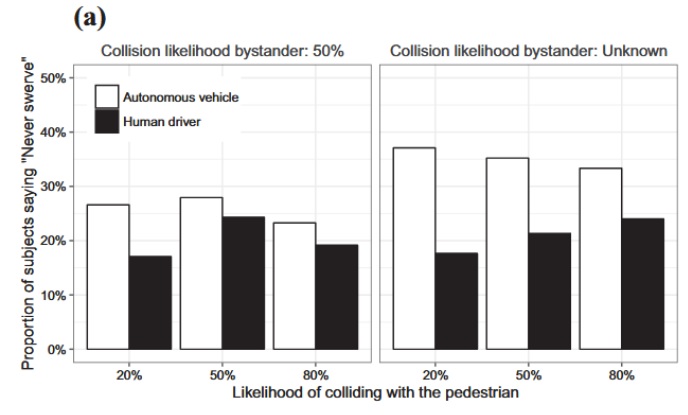}
    \caption{Proportion of participants answering "never swerve" to the third question of experiment 1.}
    \label{fig:how_q3}
\end{figure}

Similarly, fig. \ref{fig:how_q3} shows that the proportion of people saying to "never swerve" was consistently higher for autonomous cars in the third question.

In light of the foregoing, it can be concluded that subjects placed a particular weight on a \textit{default action} (putting on the brakes and staying in lane), even if this approach was not the loss-minimizing one.\\

\subsubsection{Experiment 2}

The second experiment was meant to verify whether if the foregoing findings held up with hindsight, meaning if these preferences remained when the outcome of the scenario was already known.

The described scenario follows the same pattern, though here only AVs are regarded and the estimated chances of collision were both either unknown or 50\%. In addition the outcome (accident or no accident) with a taken action by the car (staying or swerving) were known to participants.

\begin{figure}[h]
    \centering
    \includegraphics[width=.48\textwidth]{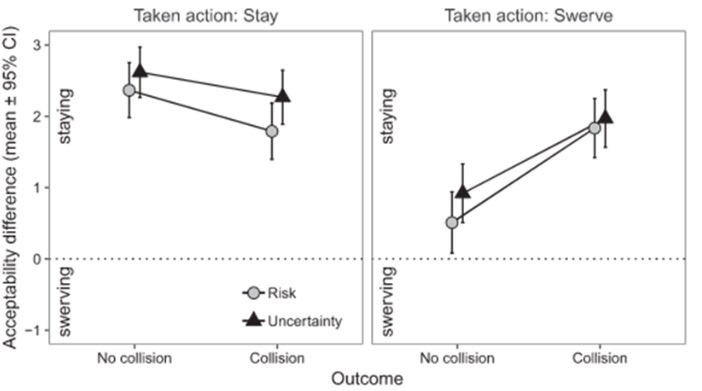}
    \caption{Moral judgement for AV actions from experiment 2.}
    \label{fig:how_e2_q2}
\end{figure}

When prompted for a moral judgement, the distribution of answers follows fig. \ref{fig:how_e2_q2}. Evidently, the decrease in moral acceptability between collision and no collision is negligible if the car stayed in lane, yet decreased sharply if it swayed, thus the \textit{hindsight effect} is significantly stronger for swaying scenarios.

The preference for default rules comes as a consequence of the underlying dichotomy between two different consequentialist rulesets:
\begin{itemize}
    \item \textbf{Simple Consequentialism:} A simple consequentialist model functions in accordance to simple utilitarian arithmetic. The moral act is the one that statistically maximizes some social utility criterion. Only the direct consequences of that one particular act are considered.
    \item \textbf{Rule Consequentialism:} In a rule consequentilist model, the rightness of an action derives from whether if the action maximizes utility in a whole class of situations governed by a rule.
\end{itemize}{}

Applying a default rule such as \textit{"putting on the brakes and staying in lane"} throughout all trolley-like scenarios increases the predictability of AVs and allows other drivers and pedestrians to adapt accordingly. If we instead desire the statistical minimization of casualties above all else, AV driving patterns become unpredictable and, as a result, untrustworthy, given that pedestrians and drivers could never foresee when a car may decide to harm them in pursuit of the greater good.

The tremendous amount of focus placed on highly unrealistic, statistically unlikely predicaments such as the trolley problem, may be a primary cause to the lack of trust, as it paints AVs in an unfamiliar, cold and calculating light.

\section{CONCLUSIONS}

The future adoption-rate of self-driving cars will entirely depend on customers' willingness to purchase them. Perceiving AVs as either dangerous, untrustworthy or immoral will cause these rates to suffer.

Disproportionate exposure to news articles and headlines on AV accidents may paint a picture of a death trap on wheels, despite all the brazen promises and claims on safety improvements. Mediating expectations may be the primary amendment to this issue by not promising infallibility and emphasizing the overall safety benefits.

Although consumer deterrence may come in consequence of extreme (bad intervention) cases, the looming threat of negative press coverage also ensures that manufacturers are kept under pressure to perform. The pedantic gaze provided by a critical press implies that AV manufacturers cannot simply shirk responsibility for their products' lapses, as they will be severely questioned about any and all shortcomings.

Furthermore, familiarizing AV driving patterns through the utilization of default actions and rules may be a quintessential step in pursuit of a trustworthy kinship between human and machine drivers.

\addtolength{\textheight}{-12cm}   




\end{document}